\documentclass[11pt]{iopart}

\usepackage{graphicx}
\expandafter\let\csname equation*\endcsname\relax

\expandafter\let\csname endequation*\endcsname\relax

\usepackage{mathtools}
\usepackage{microtype}
\usepackage[usenames,dvipsnames]{color}
\usepackage{subfigure}

\begin{document}



\title[Impact of mol. CX in SOLPS-ITER]{Investigating the impact of the molecular charge-exchange rate on detached SOLPS-ITER simulations}



\author{K. Verhaegh$^{1,*}$, A. C. Williams$^{2,*}$, D. Moulton$^1$, B. Lipschultz$^2$, B. P. Duval$^3$, O. F\'{e}vrier$^3$, A. Fil$^{1}$, J. Harrison$^1$, N. Osborne$^4$, H. Reimerdes$^3$, C. Theiler$^3$, the TCV team$^**$ and the EuroFusion MST1 team$^{***}$}
\address{$^1$ United Kingdom Atomic Energy Authority, Culham, United Kingdom} 
\address{$^2$ York Plasma Institute, University of York, United Kingdom}
\address{$^3$ Swiss Plasma Centre, \'{E}cole Polytechnique F\'{e}\'{e}rale de Lausanne, Lausanne, Switzerland}
\address{$^4$ University of Liverpool, Liverpool, United Kingdom}
\address{$^*$ These authors contributed equally}
\address{$^{**}$ See author list of "H. Reimerdes, et al. 2022 Nucl. Fusion 62 042018"}
\address{$^{***}$ See author list of "B. Labit et al 2019 Nucl. Fusion 59 086020"}

\ead{kevin.verhaegh@ukaea.uk}

\begin{abstract}

Plasma-molecular interactions generate molecular ions which react with the plasma and contribute to detachment through molecular activated recombination (MAR), reducing the ion target flux, and molecular activated dissociation (MAD), both of which create excited atoms. Hydrogenic emission from these atoms have been detected experimentally in detached TCV, JET and MAST-U deuterium plasmas. The TCV findings, however, were in disagreement with SOLPS-ITER simulations for deuterium indicating a molecular ion density ($D_2^+$) that was insufficient to lead to significant hydrogenic emission, which was attributed to underestimates of the molecular charge exchange rate ($D_2 + D^+ \rightarrow D_2^+ + D$) for deuterium (obtained by rescaling the hydrogen rates by their isotope mass).

In this work, we have performed new SOLPS-ITER simulations with the default rate setup and a modified rate setup where ion isotope mass rescaling was disabled. This increased the $D_2^+$ content by $> \times 100$. By disabling ion isotope mass rescaling: 1) the total ion sinks are more than doubled due to the inclusion of MAR; 2) the additional MAR causes the ion target flux to roll-over during detachment; 3) the total $D\alpha$ emission in the divertor increases during deep detachment by roughly a factor four; 4) the neutral atom density in the divertor is doubled due to MAD, leading to a 50\% increase in neutral pressure; 5) total hydrogenic power loss is increased by up to 60\% due to MAD. These differences result in an improved agreement between the experiment and the simulations in terms of spectroscopic measurements, ion source/sink inferences and the occurrence of an ion target flux roll-over. Extrapolating simplified scalings of divertor molecular densities (TCV \& MAST-U) to reactor-relevant densities suggests the underestimation of molecular charge exchange could strongly impact divertor physics (neutral atom density, ions sinks) and hydrogen emission (which has implications for detachment control) in deeply detached conditions, warranting further study.

\end{abstract}


\section{Introduction}
\label{ch:introduction}

Power exhaust is an important challenge in the realisation of fusion power plants as the tolerable heat flux engineering limits to the vessel walls may otherwise be greatly exceeded. Large heat flux reduction requires plasma detachment which is triggered when relatively low temperatures ($T_e < \sim 5$ eV) are obtained; these can be achieved by increasing the core density and/or inducing impurity radiation by extrinsic impurity seeding. Detachment is a state where plasma-atom and molecule interactions result in simultaneous power, particle and momentum dissipation; which is experimentally recognised by a reduction of the ion target flux. Although the incoming flux can mostly be considered atomic, the recycling of the ions from the wall can have a strong molecular component. Therefore, the relevant collisional processes include atomic processes, such as electron-impact excitation (EIE power loss), electron-impact ionisation (ion source) and electron-ion recombination (EIR ion sink); as well as molecular processes. Interactions between the plasma and the molecules can lead to momentum and energy transfer from the plasma to the molecules, exciting molecules rovibronically (i.e. rotationally, vibrationally and electronically). Vibrational excitation ($\nu$) increases the probability of creating molecular ions, in particular $D_2^+$ (through molecular charge exchange $D_2  (\nu)+ D^+ \rightarrow D_2^+ + D$) and, potentially, $D^-$ (through $e^- + D_2 (\nu) \rightarrow D_2^- \rightarrow D^- + D$). These ions can then undergo further interactions with the plasma, leading to Molecular Activated Recombination (MAR) \cite{Krasheninnikov1997} and Molecular Activated Dissociation (MAD) \cite{Krasheninnikov1997}; that result in excited atoms and, thus, radiative losses.

Recent experimental investigations have used spectroscopy and filtered camera imaging on TCV \cite{Verhaegh2017,Verhaegh2019,Verhaegh2019a,Verhaegh2021,Verhaegh2021a,Verhaegh2021b,Perek2019submitted,Perek2021,Perek2022}, MAST-Upgrade \cite{Verhaegh2022} and JET \cite{Lomanowski2020,Karhunen2022,Karhunen2022a} that register significant Balmer line emission after the break-up of molecular ions. For TCV \cite{Verhaegh2021a,Verhaegh2021b}, it was shown that MAR from molecular ions is a dominant contributor to the reduction of the ion target flux in contradiction with SOLPS-ITER simulations \cite{Verhaegh2021a,Fil2017} that did not show a roll-over of the ion target flux \cite{Fil2017,Fil2019submitted,Wensing2019,Wensing2020}. Additionally, the $D_2^+$ density in the simulations was negligible during detachment, leading to negligible MAR and a much lower simulated $D\alpha$ emission \cite{Verhaegh2021a}. That disagreement was likely caused by an underestimated molecular charge exchange rate, leading to orders of magnitude underestimations in the $D_2^+$ content in detached conditions (sections \ref{ch:molCX} and \ref{ch:rates_improvement}) when the hydrogen rates are isotope mass rescaled to deuterium \cite{Verhaegh2021b}. More recent studies, however, indicate the difference between the molecular charge exchange rate for hydrogen and deuterium should be much smaller and may be even $< 5 \%$ at 1-3 eV \cite{Reiter2018,Kukushkin2017}.

To illustrate the impact upon SOLPS-ITER simulations, converged simulation results were "post-processed" \cite{Verhaegh2021b} to evaluate the potential impact of underestimating the $D_2^+$ content. Using the simulated $D_2$ densities, the $D_2^+$ density was modelled based upon the $D_2^+/D_2$ ratio from \cite{Kukushkin2017}. This post-processing led to a better agreement with the experimental data: 1) the MAR ion sink became significant; 2) if this MAR ion sink is subtracted from the ion target flux, an ion target flux roll-over would have occurred; 3) $D_2^+$ interactions added significant Balmer line emission. However, these effects appeared 'overestimated' after post-processing.

When integrally simulated, modifying the $D_2^+$ content leads to modifications in the plasma solution (i.e. $n_e, T_e, ...$) itself, that cannot be accounted for in post-processing. Hence, in this work, we present a comparison between self-consistent SOLPS-ITER simulations for TCV (same as those used in \cite{Fil2017,Fil2019submitted}) which use: 1) the default SOLPS-ITER rate setup that employs ion isotope rescaling to the molecular charge exchange rate; 2) a modified rate setup where the ion isotope mass rescaling of the molecular charge exchange rate is disabled, see figure \ref{fig:D2pD2_rat}. For simplicity, we refer to these as the 'default' and the 'modified' setups.

We observe that including these rates self consistently leads to an ion target flux roll-over, induced by strong MAR ion sinks, and an increase in the $D\alpha$ emission - in agreement with experiment. Additionally, we find that the neutral atom content in the divertor is more than doubled through MAD, which significantly elevates the hydrogenic plasma power loss channel. 

The aim of this work is to show that molecular charge exchange can impact detachment significantly in SOLPS-ITER simulations (in agreement with the experiment) and to motivate the necessity of investigating the reaction rate set in more detail. Since re-deriving and using new molecular charge exchange rates is outside of the scope of this work, we have chosen to do this by disabling the ion isotope mass rescaling for modelling simplicity. The motivation for this (section \ref{ch:molCX}) is that there are various inaccuracies in the molecular charge exchange rates employed by SOLPS-ITER at temperatures below 2 eV. Although, there is a \emph{physics} reason for ion isotope mass rescaling (section \ref{ch:molCX}), it exacerbates the various rate problems at temperatures below 4 eV (for D) and below 6 eV (for T), respectively. Disabling ion isotope mass rescaling is not advisable as a new SOLPS-ITER default and improved rate data needs to be implemented in Eirene (section \ref{ch:rates_improvement}).

\section{Simulation setup and theory}

\begin{figure}
    \centering
    \includegraphics[width=0.4\linewidth]{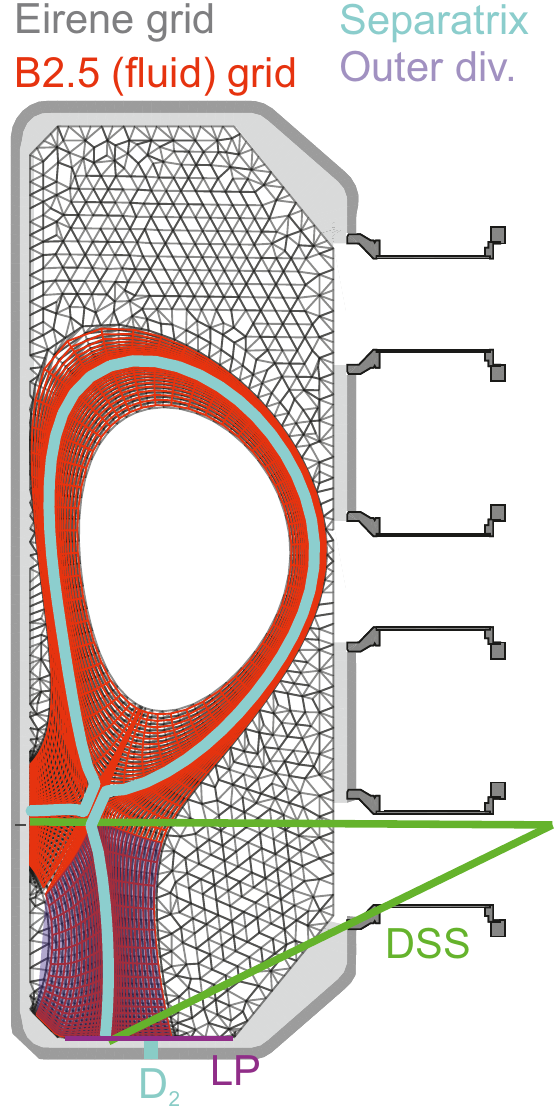}
    \caption{A visualisation of the TCV vessel geometry, fluid grid (red), Eirene grid (black/grey), spectroscopic (DSS) lines of sight (green), outer divertor fluid grid cells (blue shading), separatrix fluid grid cells (cyan), $D_2$ fuelling location (cyan) and Langmuir probe (LP) location (magenta).}
    \label{fig:Grid}
\end{figure}

The simulation setup follows the previously published work by Fil \cite{Fil2017,Fil2019submitted} \footnote{For provenance, this was re-run with a newer SOLPS-ITER version (version 3.0.7), which did not yet have neutral-neutral collisions as a default.}. The original simulations performed were interpretive simulations of TCV tokamak \cite{Coda2019,Reimerdes2022}, discharge \# 52065, which is a single null L-mode unbaffled \cite{Reimerdes2021,Raj2022} density ramp discharge that has been extensively reported in literature \cite{Verhaegh2019,Verhaegh2021b,Verhaegh2021,Perek2021,Harrison2017,Reimerdes2017}. In the simulation, the upstream density is scanned by a fuelling puff, similarly to \cite{Fil2017,Fil2019submitted}. Photon opacity, neutral-neutral collisions and drifts are not included; currents are included. \cite{Verhaegh2021b} calculated that photon opacity is expected to play a negligible role in the simulated cases. Chemical sputtering of carbon is included as a fixed fraction of the ion target flux, estimated by matching the magnitude of carbon emission signals in the divertor to experiments \cite{Fil2017,Verhaegh2019}. The power input in the simulation was matched to the experiment by defining $P_{SOL}$ as the Ohmic power minus core + SOL (above X-point) radiative losses (equivalent to \cite{Verhaegh2019}) and by dividing $P_{SOL}$ equally amongst electrons and ions. This accounts for both core radiation as well as radiation in the SOL that may occur from interactions between the SOL plasma and the main chamber wall (which may be enhanced by shoulder formation \cite{Wischmeier2005}. An illustration of the plasma grid, Eirene grid and vessel geometry used is shown in figure \ref{fig:Grid}.

The simulations are set-up to run with kinetic neutrals (e.g. Eirene) using the default Eirene rate setup. The rates used by Eirene are derived for a hydrogen plasma (for simplicity, as isotope resolved data is often not available), whereas a deuterium plasma is simulated. Rates ($<\sigma v>$)  that involve interactions with ions, depend on the kinetic velocity of the ions. Since the velocity, \emph{at the same ion temperature} is lower for heavier particles, ion isotope mass rescaling to $<\sigma v> (T_i)$ is applied by default \cite{Kotov2009}. This \emph{assumes} that there is no isotope effect in the \emph{cross-sections} (e.g. no chemical isotopical differences) - evidence for this, depending on the reaction, can be sparse. 

\subsection{Molecular charge exchange rates}
\label{ch:molCX}

It has been hypothesised \cite{Verhaegh2021b} that the $D_2^+$ content may be severely underestimated in SOLPS-ITER simulations in detachment relevant regimes. Eirene uses the effective molecular charge exchange rate tabulated in 'AMJUEL' \cite{AMJUEL} by polynomial fit coefficients. Such effective rates average the reaction rate for molecular charge exchange $<\sigma v>_{CX, H_2 (\nu)}$ per vibrational state ($\nu$) over the vibrational distribution of the molecules $f_\nu$ - equation \ref{eq:eff_rate}. 

$f_\nu$ is modelled based on the local plasma parameters assuming that transport can be neglected and mostly depends on electron impact collisions with the molecules, changing the vibrational distribution \cite{Sawada1995} depending on the electron temperature $T_e$: $f_{\nu} (T_e, ...)$. The $<\sigma v>_{CX, H_2 (\nu)}$ rate is obtained from the $\sigma_{CX, H_2 (\nu)}$ and depends on the relative velocity between $H^+$ and $H_2$. Assuming $H_2$ is near stationary, the $<\sigma v>_{CX, H_2 (\nu)}$ rate is only a function of ion temperature. Hence, the effective reaction rate is sensitive to both the ion and the electron temperatures - equation \ref{eq:eff_rate}. The effective rate fit coefficient tables used by Eirene only have a temperature sensitivity, under the assumption that $T = T_e = T_i$. 

\begin{equation}
<\sigma v>_{CX, H_2, eff} = \sum_\nu f_{\nu} (T_e, ...) <\sigma v>_{H_2 (\nu)} (T_i)
\label{eq:eff_rate}
\end{equation}

$\sigma_{CX, H_2 (\nu=0)}$ is obtained from Janev, 1987 \cite{Janev1987} based on measurements in the 1970s by Holliday, et al. \cite{Holliday1971}. These cross-sections for the vibrational ground state are then scaled to higher vibrational levels using an analytic equation ($A_\nu (\nu)$) from Greenland, 2001 \cite{Greenland2001}: $<\sigma v>_{CX, H_2 (\nu)} (T) = A_\nu (\nu) <\sigma v>_{CX, H_2 (0)} (T)$. Although $<\sigma v>_{CX, H_2 (0)} (T)$ (and thus $<\sigma v>_{CX, H_2 (\nu)} (T) = A_\nu (\nu) <\sigma v>_{CX, H_2 (0)} (T)$) decays strongly for $T<2$ eV, this does not occur for newer \emph{vibrationally resolved} calculations of $<\sigma v>_{CX, H_2 (\nu)}$ \cite{Ichihara2000} at $H_2 (\nu \geq 4)$, which contribute most to $<\sigma v>_{CX, H_2, eff}$. The simplified Greenland scaling cannot account for this, leading to order-of-magnitude underestimates of the default $<\sigma v>_{CX, H_2 (\nu)}$ Eirene rates in detachment relevant conditions. 

For deuterium and tritium, the velocity of the ion interacting with the molecule at \emph{same ion temperature} is reduced by the isotope mass. Assuming that the $\sigma_{CX, H_2 (\nu)}$ cross-sections are the same for all hydrogen isotopes, the vibrationally resolved rates can be 'ion isotope mass rescaled' from hydrogen to deuterium: $<\sigma v>_{CX, D_2 (\nu)} (T_i) = <\sigma v>_{CX, H_2 (\nu)} (T_i/2)$ (see A.4.2 in \cite{Kotov2007}). \emph{However}, since Eirene only knows about the \emph{effective rate} (for a vibrationally unresolved setup), the \emph{total effective rate} is ion isotope mass rescaled, inadvertently also rescaling the electron temperature dependency of the vibrational distribution incorrectly (equation \ref{eq:eirene_rate}). 

\begin{equation}
\begin{split}
 <\sigma v>_{CX, D_2, eff, Eirene} (T) &= <\sigma v>_{CX, H_2, eff} (T/2) \\
 &= \sum_\nu f_{\nu} (T/2, ...) <\sigma v>_{CX, H_2(\nu)} (T/2, ...) \\
  <\sigma v>_{CX, D_2, eff, correct} (T) &= \sum_\nu f_{\nu} (T, ...) <\sigma v>_{CX, H_2(\nu)} (T/2, ...)
  \end{split}
\label{eq:eirene_rate}
\end{equation}

\begin{figure}
    \centering
    \includegraphics[width=0.8\linewidth]{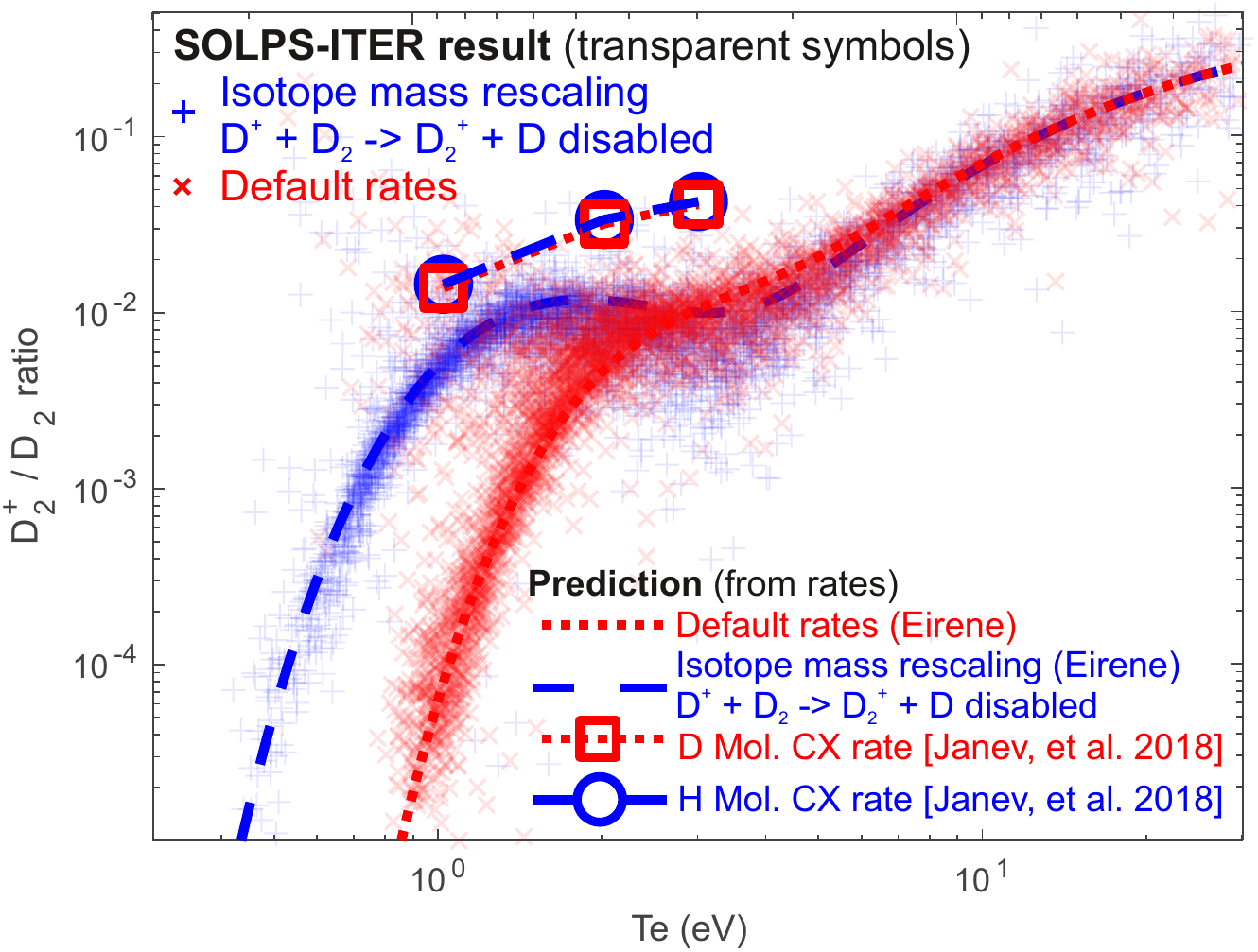}
    \caption{The $D_2^+/D_2$ ratio obtained from SOLPS simulations (transparent symbols), over the entire simulation domain, as a function of the electron temperature for both the default SOLPS-ITER setup as well as the setup where mass rescaling is disabled for molecular charge exchange. The expected $D_2^+/D_2$ ratio with and without isotope mass rescaling of the molecular charge exchange rate are also shown, modelled as the sum of the creation rates of $D_2^+$ (from $D_2$) divided by the sum of the destruction rates of $D_2^+$. Additional $H_2^+/H_2$ and $D_2^+/D_2$ ratios are shown ('H, D mol CX rate [Janev, et al. 2018]') using the molecular charge-exchange rate from \cite{Reiter2018} in combination with reaction rates from \cite{AMJUEL,Reiter2008} for the remaining reactions. }
    \label{fig:D2pD2_rat}
\end{figure}

Since $<\sigma v>_{CX, H_2, eff} (T)$ decreases with decreasing $T$ \footnote{The $f_\nu (T_e, ...)$ used for the derivation of the effective rates in Eirene is not fully documented}, this rescaling greatly reduces the effective molecular charge exchange rate in detachment relevant conditions for deuterium (a factor $\sim 100$) and tritium. This is shown in figure \ref{fig:D2pD2_rat}, where the $D_2^+/D_2$ ratios obtained from SOLPS-ITER simulations with and without isotope mass rescaling as well as the theoretically expected  $D_2^+/D_2$ ratio based on a no-transport model (e.g. balancing the $D_2^+$ creation/destruction rates) are compared. It is these conditions in which molecular density greatly increases for decreasing temperatures \cite{Stangeby2018, Verhaegh2021a,Verhaegh2022}. 

In contrast, using a more accurate analysis of the molecular charge exchange rate \cite{Reiter2018} for $T = 1, 2, 3$ eV shows a negligible ($\sim 5$ \%) isotope dependency \footnote{However, this is based on a simplified Boltzmann relation for $f_\nu$, which is different from $f_\nu (T_e, ...)$ used for deriving the effective rates in Eirene.}, where three issues discussed above were resolved: 1) $<\sigma v>_\nu$ rates were obtained from \cite{Ichihara2000}, using full vibrationally resolved simulations, rather than applying the simplified Greenland rescaling; 2) rates specifically derived for $H$ and $D$ were utilised, accounting for chemical isotope differences; 3) the lower velocity of the heavier hydrogen isotopes was correctly accounted for only in $<\sigma v>_\nu (T_i, ...)$ and not in the vibrational distribution ($f_\nu (T_e, ...)$. The ratios obtained with these rates (figure \ref{fig:D2pD2_rat}), for both $H_2^+/H_2$ and $D_2^+/D_2$, are in closer agreement to the AMJUEL effective $H_2^+/H_2$ ratio; motivating our choice for disabling ion isotope mass rescaling in the 'modified' setup.



\section{Simulation results}
\label{ch:results}

\begin{figure}
    \centering
    \includegraphics[width=\linewidth]{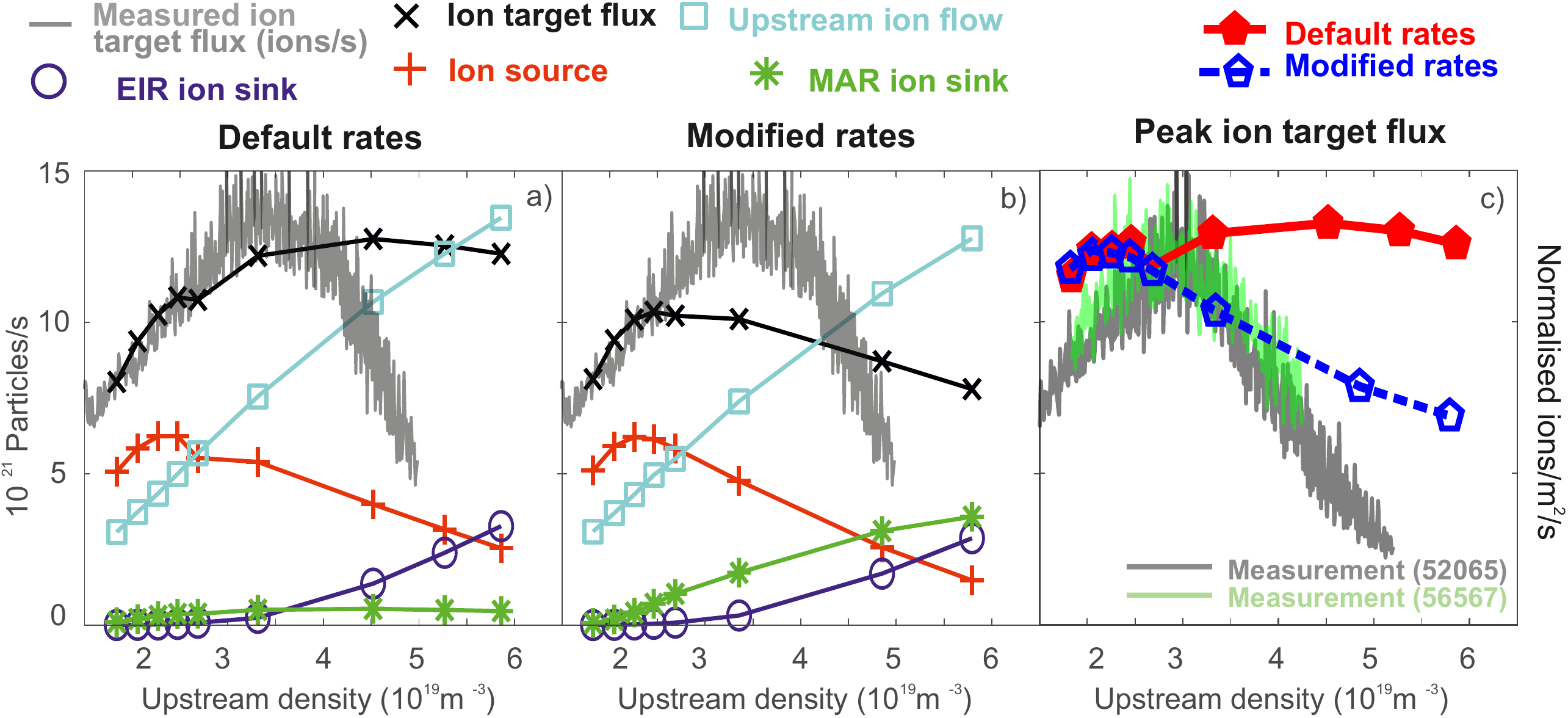}
    \caption{Obtained particle balance in terms of integrated outer divertor ion target flux; total outer divertor integrated ion sources/sinks are shown, together with the net ion flow from outside the divertor region towards the outer divertor target for a) the default SOLPS-ITER simulation setup; b) a SOLPS-ITER simulation setup where ion isotope mass rescaling for molecular charge exchange was disabled. c) A comparison of the peak ion target flux measurement (normalised ions/$m^2$/s) for \# 52065 and a repeat \# 56567 (see figure \ref{fig:exp_compa}) to the simulated peak ion target flux for the default and modified rate setup.}
    \label{fig:PartDaBal}
\end{figure}

The obtained outer target ion fluxes, together with the total outer divertor ion source (atomic ionisation and molecular activated ionisation arising from $D_2$ and $D_2^+$), ion sinks (electron-ion recombination and molecular activated recombination arising from $D_2^+$) and net ion flow into the divertor is shown in figure \ref{fig:PartDaBal} during a core density ramp, together with the experimentally measured ion target particle flux (assuming a 15 \% uncertainty \cite{Fevrier2018,Oliveira2019}) for \# 52065. The upstream density was measured combining Thomson scattering measurements with the equilibrium reconstruction of the separatrix and assuming a spatial uncertainty of $\pm 2.2$ cm (e.g. the Thomson resolution), resulting in an upstream density uncertainty of $\pm 0.5 \times 10^{19} m^{-3}$ before the ion target flux roll-over and $\pm 1.5 \times 10^{19} m^{-3}$ afterwards. Analogous to \cite{Verhaegh2019a}, to convert the time in the discharge to upstream density, a linear fit of the upstream density as function of time \emph{from before the ion target flux roll-over} is performed. \footnote{This is extrapolated to after the roll-over, because the upstream density measurements after the ion target flux roll-over are highly uncertain and evidence of saturation of the upstream density during core density ramps is occasionally found during detachment on TCV \cite{Verhaegh2019a}.}. We will now compare the 'default' and 'modified' setups, where \emph{only} the  ion isotope mass rescaling for molecular charge exchange has been disabled.

In both the 'default' and 'modified' setups, we observe a movement of the divertor ionisation source towards the X-point as the upstream density is increased, reducing the total outer divertor integrated ion source, whilst electron-ion recombination remains negligible, in agreement with experimental observations \cite{Verhaegh2019}. As the ionisation moves upstream, the ion flux from upstream of the divertor towards the outer target is increased, replenishing the loss of divertor ionisation, in agreement with measurements by a reciprocating divertor probe array \cite{Oliveira2022}, as well as upstream spectroscopic measurements \cite{Verhaegh2021}. In the 'default' setup, this prevents the roll-over of the ion target flux as the upstream density is increased. This is in contrast to the experiment, where a clear roll-over is observed as indicated in figure \ref{fig:PartDaBal} \cite{Reimerdes2017,Harrison2017,Verhaegh2017,Verhaegh2019,Fil2017}. The absence of an ion target flux roll-over in plasma-edge simulations, even though other detachment markers are clearly obtained in the simulations (movement of the ionisation source; appearance of $T_e \sim 1$ eV temperatures; volumetric momentum losses; ...), is a general observation for TCV density ramp SOLPS-ITER simulations \cite{Wensing2019,Wensing2020,Smolders,Fil2017,Fil2019submitted}. 

For the 'modified' setup, the obtained particle balance is similar to the default SOLPS-ITER setup in the attached phase. This is unsurprising, as molecular charge exchange only becomes an important source of $D_2^+$ at detachment relevant temperatures of 1-3 eV, before which the $D_2^+$ creation is dominated by $D_2$ ionisation and the isotope ion mass rescaling of molecular charge exchange has a negligible impact on the $D_2^+/D_2$ ratio. However, as the core density increases and detachment relevant temperatures are achieved, a clear increase of MAR ion sinks, together with a roll-over of the ion target flux is now observed; in contrast to the cases which used the default reaction rates. This is attributed to the additional ion sinks obtained from MAR ($D_2 + D^+ \rightarrow D_2^+ + D$ followed by $e^- + D_2^+ \rightarrow D + D$), which leads to a reduction of the ion target flux.

We find that in both cases, the contribution of MAI to the total ion source is minor: throughout the density scan it starts at 16-18 \% dropping to 5-10 \% of the total ion source during detachment for both SOLPS-ITER rate setups. The ionisation source shown in figure \ref{fig:PartDaBal} includes both electron-impact ionisation as well as MAI.

The total ionisation source however, at the same upstream density, is smaller for the 'modified' setup  after detachment - e.g. after MAR and $D\alpha$ emission arising from plasma-molecular interactions starts to become important. The ionisation front is further removed from the target, after the detachment onset, for the 'modified' setup with lower target temperatures: more 'deeply detached' scenarios (in terms of ionisation front displacement, and target temperature) are obtained. This is likely related to the additional ion sinks and hydrogenic power losses associated with  Molecular Activated Dissociation (MAD), explained in section \ref{ch:MAD}. 

Since the modifications to the molecular charge exchange rate only play a role at low temperatures ($T_e<3$ eV), it only plays a role below the ionisation region. The 'modified' setup simulations being more deeply detached with both lower target temperatures \emph{and a larger ionisation front displacement} (for the same upstream density) suggests that a process that occurs \emph{downstream} of the ionisation region can impact the ionisation region location. Although not the main point of the paper, this suggests that the divertor processes are 'self-regulating', converging to an integrated, self-consistent solution that may appear non-local: downstream processes can influence upstream processes. 

Although the ion target flux roll-over is recovered in SOLPS-ITER with modified simulations, the roll-over is more pronounced in the experiment (figure \ref{fig:PartDaBal}). The difference in roll-over point of the upstream density between experiment and simulation is within experimental uncertainty ($\pm 1.5 \times 10^{19} m^{-3}$). Upstream pressure losses and reductions of the upstream density can be measured experimentally during detachment on TCV \cite{Fevrier2019submitted,Verhaegh2019a} and have been observed for \# 56567 - a repeat of \# 52065, which are not reproduced in SOLPS-ITER \cite{Verhaegh2019a}. Such behaviour could cause a stronger roll-over in the simulations than observed and complicate the usage of the upstream pressure as an ordering parameter. 

In addition, there may be processes, other than molecular charge exchange, that may result in discrepancies between experiment and simulations. It was hypothesised in \cite{Wischmeier2005} that enhanced erosion of carbon from the main chamber wall through enhanced perpendicular transport at higher densities may result in additional impurities that strengthen detachment on TCV, which could also cause higher ion target flux losses in the experiment than simulated. However, investigating this is outside of the scope of this work and requires additional diagnostics, such as filtered imaging of the main plasma to diagnose carbon erosion \cite{Perek2022}, which should be addressed in the future. 

Neutrals can escape relatively easily in the simulated open TCV divertor, which can get ionised in the scrape-off-layer. This results in a flow of ions from upstream towards the target, which is the dominant contributor to the ion target flux for upstream densities higher than $3 \times 10^{19} m^{-3}$ (figure \ref{fig:PartDaBal} a, b). Therefore, an agreement between the measured and simulated ion target flux depends on whether the ion flow from upstream is simulated correctly, which can depend on drifts that have not been included in this work\cite{Wensing2020}. Therefore, the peak ion target flux is compared against the measurements in addition in figure \ref{fig:PartDaBal} c. The peak ion target flux saturates at a constant value for the default rates, but rolls-over for the modified rates - although not as steeply as in the experiment.

\subsection{Comparison of SOLPS-ITER results to ion source \& sink measurements}

\begin{figure}
    \centering
    \includegraphics[width=\linewidth]{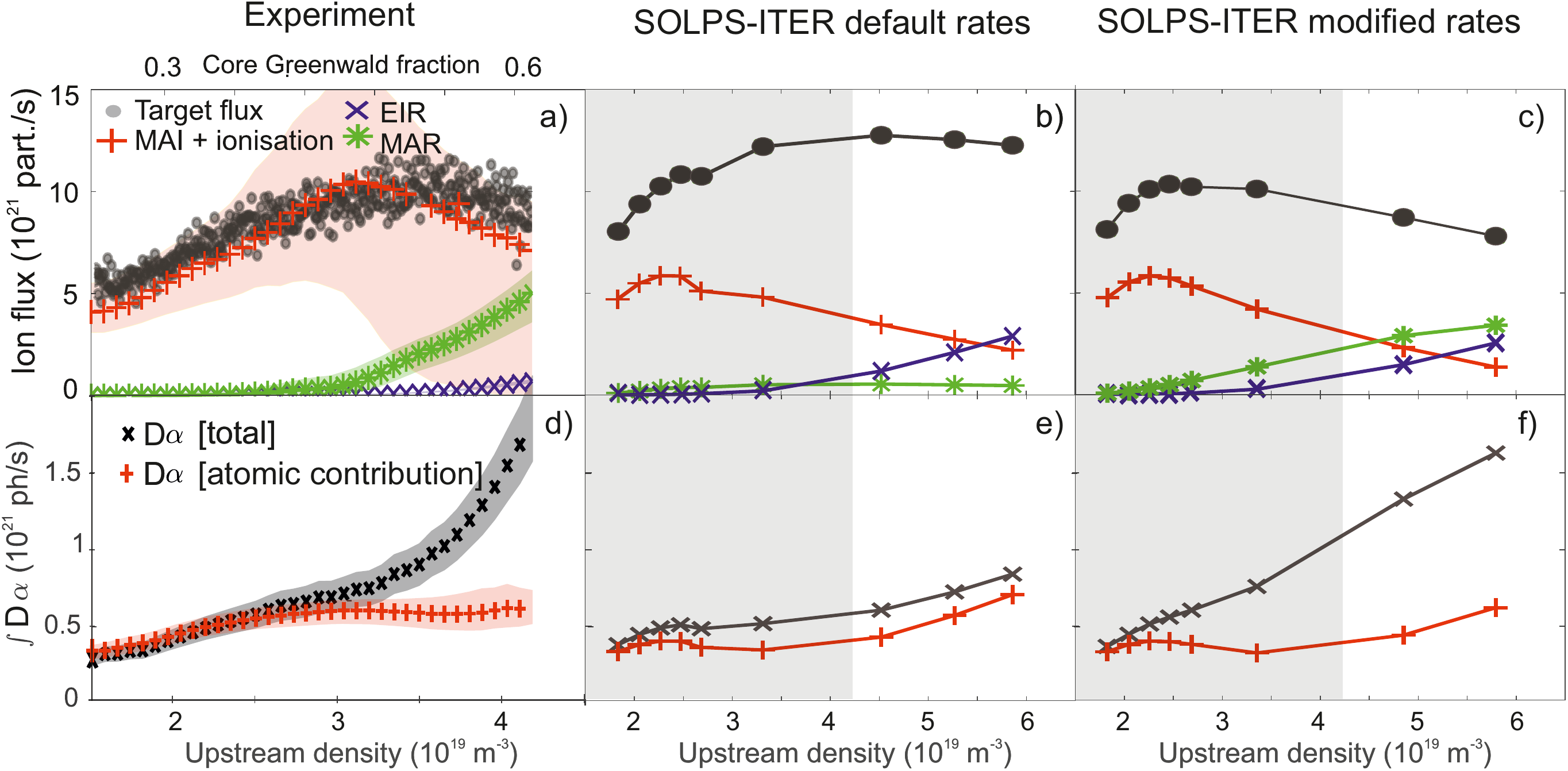}
    \caption{Obtained particle balance (top figure) in terms of integrated outer divertor ion target flux; and the ion sources/sinks obtained in view of the DSS diagnostic. Measured total $D\alpha$ emission (bottom figure) in the outer divertor, captured in between the DSS lines of sight, and its inferred contribution arising from EIE and EIR (e.g. 'atomic' interactions) and from plasma-molecular interactions. These results are shown for simulations where ion isotope mass rescaling was disabled for molecular charge exchange, which are compared against the experimental observations of outer the ion target flux (Langmuir probes), the total $D\alpha$ emission and spectroscopic inferences of the divertor ion source, ion sinks (MAR \& EIR) as well as the separation of $D\alpha$ in atomic (e.g. EIE \& EIR) and 'molecular' (associated with $D_2$, $D_2^+$ \& $D_2^- \rightarrow D^- + D$) components. The experimental data is expressed in terms of both the core Greenwald fraction and upstream density (obtained analogously to figure \ref{fig:PartDaBal}). The scanned upstream density region (indicated by the grey shading in b,c,e,f) is smaller in the experiment than in the simulations.}
    \label{fig:exp_compa}
\end{figure}

The simulated discharge, \# 52065, was repeated several times to perform a detailed spectroscopic characterisation (\# 56567 and repeats) \footnote{It should be noted that the ion target flux roll-over is less clear in \# 56567 than \# 52065 as significantly lower core densities were obtained before the plasma disrupted. The measured upstream density saturated at $3.3\times10^{19} m^{-3}$ during the density ramp, although extrapolated upstream density scalings from the attached phase reach $4.2\times10^{19} m^{-3}$ at the end of the discharge (compared to $5\times10^{19} m^{-3}$ for \# 52065). However, in both \# 52065 and \# 56567 a clear roll-over of the peak ion target flux is found (figure \ref{fig:PartDaBal} c), which is only reproduced by the modified rate setup}. This facilitated a detailed spectroscopic characterisation of the ion sources and sinks in the lower divertor \cite{Verhaegh2017,Verhaegh2019,Verhaegh2019a,Verhaegh2021,Verhaegh2021a, Verhaegh2021b} and separating the measured $D\alpha$ emission into its various emission contributions \cite{Verhaegh2021,Verhaegh2021a,Verhaegh2021b}. That result is shown in figure \ref{fig:exp_compa}, together with the obtained simulation results where the ion sources/sinks have been integrated over the spectroscopic lines of sight. This previously reported experimental analysis \cite{Verhaegh2021,Verhaegh2021a,Verhaegh2021b} indicates 1) significant ion sinks through MAR, dominating over EIR ion sinks; 2) a strong increase of $D\alpha$ due to emission from excited atoms after plasma-molecular interactions. Both these aspects are absent in the 'default' setup simulations, but are present in the 'modified' setup with similar magnitudes as in the experiment.

\subsection{The impact of molecular charge exchange on neutral atom content and hydrogenic power losses}
\label{ch:MAD}

\begin{figure}
    \centering
    \includegraphics[width=\linewidth]{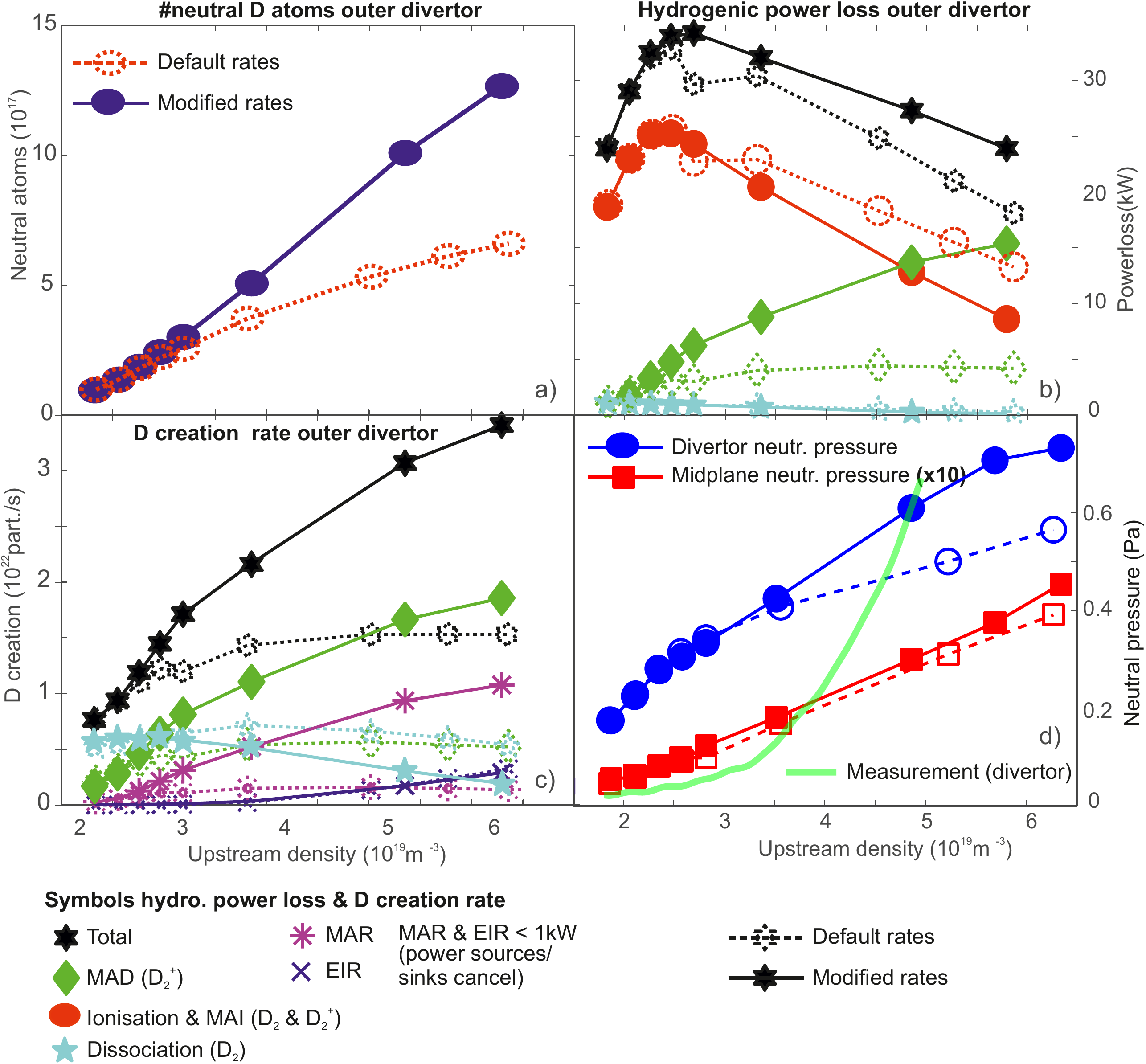}
    \caption{The impact of increased MAD, due to the modified reaction rates, on hydrogen atom content, hydrogenic power losses and neutral pressure. a) Evolution of the total neutral atom content (excluding molecules) as function of upstream density for the default and modified rates. b) Evolution of hydrogenic power loss processes for the default and modified rates, including: ionisation power loss (sum of radiative power loss due to excitation collisions preceding ionisation and the potential energy, $\epsilon=$ 13.6 eV, spent on ionisation); power losses associated with electron-impact dissociation and associated with MAD. The net power loss associated with MAR \& EIR has been estimated to be below 1 kW. c) Volumetric neutral atom creation source, integrated over the outer divertor, from MAR, MAD, EIR and electron-impact dissociation for the default and modified rate setup; d) neutral pressures obtained using a synthetic baratron using the simplified model from \cite{Wischmeier2005,Wensing2019} for the divertor and midplane (multiplied by 10), together with the experimental data for \# 52065 for the divertor baratron. Relevant symbols for b \& c are listed below the figures and in all figures dashed lines with open symbols represent the default rate simulations and solid lines with filled symbols represent the modified rate simulations.}
    \label{fig:NeutralSource}
\end{figure}

Disabling the ion isotope mass rescaling for the molecular charge exchange reaction not only has a strong impact on the divertor particle balance and hydrogenic emission processes, but also on the neutral atomic content. The total neutral D atom content (excluding molecules) in the outer divertor is increased by more than a factor of two for identical upstream densities with the 'modified' compared to the 'default' setup (figure \ref{fig:NeutralSource} a), at the deepest detached phases. Even when the neutral atom content between the two simulation setups is compared as a function of the target temperature, the neutral atom content is enhanced by more than 50 \% for the modified rate setup (not shown). However, since the electron density decays in the modified rate setup, the total nuclei content (e.g. total amount of $D$ particles considering ions, atoms and molecules) remains similar (within 5 \%) for both simulation setups as function of upstream density.

This increase in neutral atom content can be explained by the additional neutral atoms created through MAR \& MAD by the modified rate. The strength of volumetric processes generating neutral atoms (e.g. EIR, MAR, MAD and electron-impact dissociation) is compared between the default and modified rates in figure \ref{fig:NeutralSource} c). This shows that the volumetric creation of neutral atoms is significantly enhanced in the modified rate setup, where the neutral atom creation continuously increases as one goes into deeper detached regimes, due to MAR \& MAD. MAR \& MAD provide additional dissociation  processes that are significant below 5 eV, by which time the electron-impact dissociation cross-sections are strongly reduced, but the molecular content is strongly increased \cite{Stangeby2018,Verhaegh2022}. Therefore, MAR \& MAD can be stronger neutral atom creation processes than electron-impact dissociation \& EIR, particularly in detached regimes.

Analogous to power losses due to ionisation, there are potential (plasma) energy losses associated with molecular dissociation. The additional dissociation mechanisms through MAD after the detachment onset result in a significant increase in the total effective hydrogenic (plasma) power losses (figure \ref{fig:NeutralSource} b). The hydrogenic power losses associated with plasma-molecular interactions are due to the dissociative energy losses to the plasma channel since the radiative losses and potential energy gains from MAR roughly cancel \cite{Verhaegh2021b} \footnote{The dissociation cost itself is a power loss only from the plasma channel: this potential energy can be released back to the target as hydrogen atoms re-associate into molecules. Therefore, this may not result in target heat load reductions, unless the distance between the dissociating area and the target is significant such that the higher energy atom population has room to dissipate radially.}. Although the total hydrogenic power loss \emph{at the same upstream density} is only 20 \% higher for the modified rate at the same upstream density, the modified rate simulations are more deeply detached. When comparing hydrogenic power loss rates at the same level of detachment (e.g. similar $T_t$ and ionisation front positions), the total hydrogenic power loss can be 60 \% higher for the modified rate setup. 

Using a synthetic diagnostic pressure gauge ('baratron') setup \cite{Wensing2020,Wischmeier2005,Verhaegh2018}, the divertor neutral pressure has been calculated and compared against the experiment (figure \ref{fig:NeutralSource} d). We find that the divertor neutral pressure starts to diverge between the modified rates and the default rates at the detachment onset. At the deepest levels of detachment, the divertor neutral pressure is increased by up to 50 \% when the modified rates are used (at the same upstream density). Experimentally, a strong increase in the divertor neutral pressure is observed after detachment, with divertor pressures of up to 0.6 Pa at the deepest levels of detachment (upstream density of $n_e = 4.5 [3.4 - 6] \times 10^{19} m^{-3}$). This agrees with the 'modified' setup simulations, but only at the deepest levels of detachment (0.6 [0.47 - 0.74] Pa). An increase of the neutral pressure at the midplane has also been observed for the 'modified' rate setup, which could arise due to neutral atoms, generated from MAD, transporting upstream due to the open nature of the TCV divertor. The divertor pressure obtained during the attached phase is overestimated (by a factor $\sim 4$) in the simulations for both the modified and default rate setup, in agreement with previous TCV results \cite{Wensing2020}. The origin of this discrepancy is unknown and is inconsistent with the agreement of the Balmer line emission and the inferred ionisation sources between the experiment and the simulation \cite{Verhaegh2019}. The increase of the neutral pressure during detachment is stronger in the experiment than simulated.

Although the neutral atom density is doubled in the divertor during deep detachment with the 'modified' rate setup, the volumetric momentum losses are similar between the 'default' and 'modified' rate setups: the pressure loss at the separatrix is $\sim 3 \%$ higher for the 'modified' rate setup below 1.5 eV (where the pressure loss is 88\% or more for both setups). Decomposing the various momentum loss mechanisms \cite{Moulton2017,Park2018}, we find that - for both rate setups -  momentum loss from plasma-atom interactions is dominant for upstream densities below $3\times10^{19} m^{-3}$ and momentum loss from plasma-molecular collisions (not $D_2^+$) are dominant at higher densities. Elastic collisions of ions with molecules have been found to remove significant momentum in SOLPS-ITER simulations \cite{Moulton2017,Myatra}, which may explain the correlation between the increase in the $D_2$ density and the appearance of strong momentum losses at low temperature \cite{Stangeby2018}. Momentum losses from plasma-atom interactions are increased by up to 120\% in the 'modified' rate setup when the target temperature reaches 1.2 eV and lower, which is attributed to the higher neutral atom densities due to MAD. Simultaneously, momentum losses associated with plasma-molecular interactions are reduced in the modified rate setup by up to 25\%; which almost, but not fully, counteracts the increased momentum losses from plasma-atom interactions. 

\section{Implications, relevance and accuracy of our findings and future pathways}
\label{ch:relevance}
\setcounter{footnote}{0}
Increasing the $D_2^+$ content through the 'modified' rate setup in our work results in: 1) increased neutral atom sources through MAD and associated hydrogenic power losses; 2) significantly enhanced hydrogenic atomic line emission from excited atoms after plasma-molecular interactions; and 3) additional ion sinks through MAR, resulting in the ion target flux roll-over. Such interactions start becoming significant at detachment onset with a spatial preference towards the target side of the ionisation region where the molecular density builds up. The TCV simulations, consistent with TCV \cite{Verhaegh2021,Verhaegh2021a,Verhaegh2021b,Perek2022} and MAST-U \cite{Verhaegh2022} experimental results, indicate that plasma-molecular chemistry involving molecular charge exchange generating $D_2^+$ and associated MAD neutral atom sources, MAR ion sinks and atomic hydrogen emission: 1) starts to occur from the detachment onset onwards as the ionisation and electron-impact dissociation regions detach from the target; 2) increase in magnitude as higher molecular densities are obtained below the ionisation region when $T_e$ drops below 3 eV \cite{Stangeby2018}. 

Neutral baffling may play a strong role in this point 2), which was brought forward as an explanation for why plasma-molecular effects plays a much more significant role in the MAST Upgrade Super-X divertor than the TCV open divertor (experiments from 2016 before baffles were present) \cite{Verhaegh2022}. Based on this, molecular charge exchange could have a stronger impact on simulations for more tightly baffled divertors, which is supported by our analysis in section \ref{ch:simple_MAR_model}. In such conditions, the additional neutrals generated in the divertor baffle would not be able to move towards upstream. This could increase the neutral density and neutral pressure (figure \ref{fig:NeutralSource} d) beyond the impact observed in this work and would reduce the impact of molecular charge exchange on the upstream neutral pressure. Those additional divertor neutral atoms could also strengthen energy and momentum transfer from plasma-neutral atom elastic collisions.

\subsection{Could molecular ions play a role in reactors during detachment?}

An important question is whether such interactions are also relevant for reactors. Answering that question requires further investigation, including further studies on the applicable molecular charge exchange rates as well as applying those to a range of reactor conditions, which is outside the scope of this work. Although there are large uncertainties regarding the molecular charge exchange rates, the molecular vibrational distribution (that determines these effective rates to a large extent) and their applicability to reactors, signatures of the impact of molecular ions on the hydrogen emission are being observed in JET with the ITER-like wall \cite{Lomanowski2020, Karhunen2022,Karhunen2022a} during deep detachment. 

The ion isotope mass rescaling has been applied correctly (e.g. using $< \sigma v>_{CX, D_2, eff, correct}$ as opposed to $< \sigma v>_{CX, D_2, eff, Eirene}$ - equation \ref{eq:eirene_rate}) for a limited set of SOLPS-ITER simulations in \cite{Kukushkin2017}. Although MAR was a significant ion sink with the modified rates, it led to a reduction of EIR as the electron density in the simulation was reduced: the target profiles obtained by the SOLPS-ITER simulations were similar. This was hypothesised to be associated with increased power limitation of the ionisation source due to energy losses associated with MAR and MAD \cite{Kukushkin2017} \footnote{The underlying vibrationally resolved cross-sections used in the Eirene rates are likely underestimated at low temperatures, as shown in section \ref{ch:rates_improvement}, which can result in a significant under-prediction of molecular charge exchange even if ion isotope mass rescaling is applied correctly.}.

However, for molecular ions to potentially play a role in reactors, there must be a sufficiently high number of molecules interacting with the plasma, which requires 1) a sufficiently large distance between the downstream end of the ionisation region and the target; 2) sufficiently high molecular density, which requires low target temperatures ($< 3$ eV) and, potentially, neutral baffling. This implies that molecular ions likely could play a stronger role in reactor scenarios that feature divertor designs \& operation where the ionisation region is sufficiently detached from the target, with $T_e$ dropping to 1-3 eV below the ionisation region, to have a significant MAR rate. Although this may be feasible in current designs of reactor-class devices with conventional divertors, such as ITER and (potentially) DEMO, this may be more achievable in alternative divertor concept designs \cite{Militello2021,Kuang2020,Wigram2019} and, potentially, X-point radiator designs \cite{Bernert2020,Pan2022,Stroth2022}.  Existing plasma-edge simulations of reactors could be post-processed to assess, in a simple way, whether molecular charge exchange can play a role in reactors \cite{Verhaegh2021, Verhaegh2021a, Verhaegh2021b}. This allows estimating whether a modified rate could impact hydrogen emission, MAR ion sinks and MAD neutral atom sources significantly \emph{in a non-self consistent way}. This can only be used to map out whether molecular ions could potentially play a strong role in a simulation, self-consistent simulations are then required to investigate the precise impact of molecular ions.

The impact of transport and plasma-wall interactions on the vibrational distribution (section \ref{ch:vibr_distr}) can be different in reactors than in devices like TCV and MAST-U. Differences in power and density result in a shortening of the mean free path in reactors, making transport of vibrationally excited molecules less likely (although it can still be significant in the low temperature region below the ionisation front). Differences in wall material (metal for reactors, carbon for TCV and MAST-U) can impact the initial vibrational distribution of the molecules coming off the wall \cite{Wischmeier2005}; as well as the reflection of atoms from the wall (in contrast to the adsorption of atoms to the wall, after which re-association occurs and molecules are released back into the plasma), which may relatively reduce the molecular density. 

\subsubsection{Simplified MAR rate modelling}
\label{ch:simple_MAR_model}

\begin{figure}
    \centering
    \includegraphics[width=\linewidth]{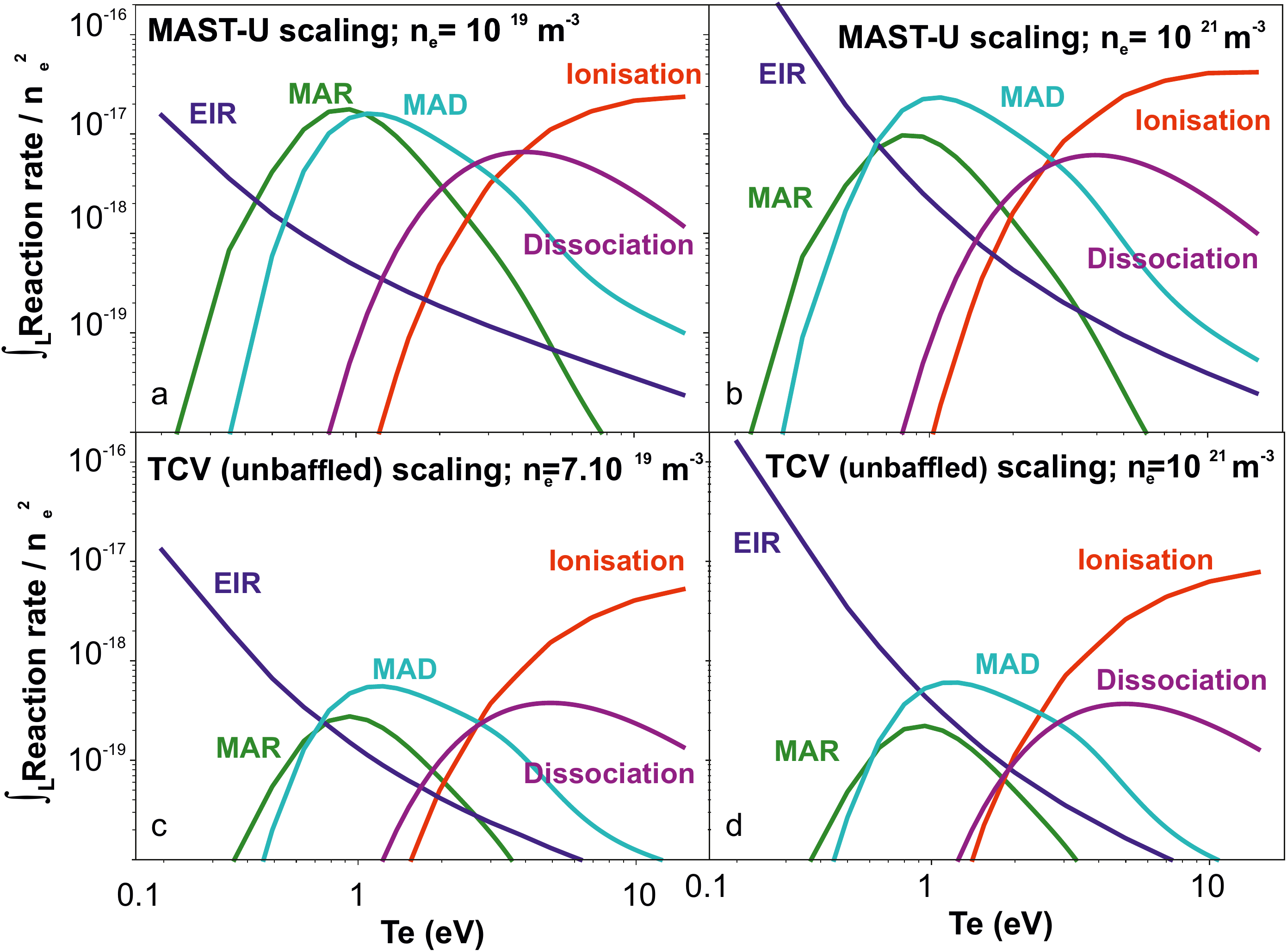}
    \caption{Comparison of the expected chordally integrated reaction rate scalings as function of $T_e$ for the TCV (c, d) \cite{Fil2017} (unbaffled) and MAST Upgrade Super-X (a, b) \cite{Myatra} divertor, based on SOLPS-ITER scalings \cite{Verhaegh2022} at a characteristic density (a,c) and extrapolated scalings to a reactor-relevant density ($n_e = 10^{21} m^{-3}$) (b,d).}
    \label{fig:ReacRate}
\end{figure}

One argument as to why plasma-molecular interactions may play a relatively weaker role for reactors is that reactors will operate at significantly higher electron densities ($n_e \sim 10^{21} m^{-3}$). Since the EIR source scales $\propto n_e^{2-3}$ for $T_e < 1.5$ eV, it would be expected that the relative role of EIR increases for reactor-relevant conditions at low temperatures; potentially reducing the relative impact of MAR as a neutral atom source and hydrogen ion sink. To investigate this further, we have obtained scalings from previous SOLPS-ITER simulations of both MAST Upgrade \cite{Myatra} and TCV \cite{Fil2017} (open divertor - without baffles) for the evolution of the spectroscopic line-of-sight integrated neutral atom and neutral molecular density relative to the electron density as function of $T_e$ \cite{Verhaegh2022} using all of the divertor spectroscopic lines of sight for both TCV and MAST-U. Using a simplified model for the rates \cite{Verhaegh2022}, together with the $D_2^+/D_2$ ratios with the 'modified' rate setup (figure \ref{fig:D2pD2_rat}), the evolution of the various atomic and molecular rates can be calculated as function of $T_e$ for a fixed $n_e$. That result is shown in figure \ref{fig:ReacRate} using the characteristic TCV \& MAST-U electron densities ($n_e = 7 \times 10^{19} m^{-3}$ and $n_e = 10^{19} m^{_3}$, respectively) as well as reactor-relevant $n_e=10^{21} m^{-3}$ extrapolations. 

The evolution of the reaction rates (figure \ref{fig:ReacRate}) indeed indicates that, when the SOLPS-ITER scalings for TCV are used, EIR becomes more important than MAR at reactor-relevant densities. However, MAD \footnote{Note that a single MAR/MAD reaction can result in the creation of 3-2 neutral atoms; whereas EIR only results in 1 neutral atom.} still remains important between $\sim 0.7$ and $\sim 3$ eV for the TCV SOLPS-ITER scalings. Therefore, even at reactor-relevant densities, MAD could be a dominant neutral generation rate - even when scalings for an open divertor are extrapolated. Strikingly, when the MAST Upgrade derived SOLPS-ITER scalings are used, we find that MAR+MAD can be dominant between 0.5 and 4 eV for reactor-relevant densities. The difference between this result and that from the unbaffled TCV scaling is associated with the higher molecular content in MAST-U, likely due to its tight baffling \footnote{The neutral leakage in current, lower power, MAST-U experiments may be more significant than in the higher power predictive simulations \cite{Myatra} figure \ref{fig:ReacRate} is based on, based on the monitored electron-impact excitation emission region \cite{Verhaegh2022}.}. This shows that indeed, if one can have a significantly high molecular density below the ionisation region, MAR can remain important despite EIR being strongly elevated at reactor-relevant densities. Therefore, molecular charge exchange may have a bigger impact in simulations for tokamaks with tightly baffled divertors than presented in this paper. 

Using effective Eirene rates without ion isotope mass rescaling, the simplified model results in figure \ref{fig:ReacRate} predict that plasma-molecular interactions involving molecular ions are negligible at very low temperatures ($T_e \ll 0.5$ eV), in contrast with results from MAST Upgrade experiments in the Super-X divertor, where such interactions are still experimentally inferred at $T_e \ll 0.5$ eV \cite{Verhaegh2022}. This mismatch is likely caused by the underestimated charge exchange cross-sections at low temperatures in Eirene (sections \ref{ch:molCX} \& \ref{ch:rates_improvement}). 

The applied scalings, derived from SOLPS-ITER, will be different in reactor-relevant conditions: our result does not predict that MAR \& MAD will be important for reactors. It serves as additional motivation as to why molecular ions leading to MAR \& MAD require further study in reactor relevant regimes.

\subsection{Impact on diagnostic design, analysis and real-time detachment control strategies}

MAR \& MAD not only have an impact on the divertor physics, but also result in a significant content of excited hydrogen atoms and thus hydrogen Balmer line emission as was shown in literature \cite{Hollmann2006,Karhunen2022a,Karhunen2022,Lomanowski2020,Verhaegh2021,Verhaegh2021b,Verhaegh2021a,Verhaegh2022} and is indicated by the increase in $D\alpha$ emission during detachment as shown in figure \ref{fig:PartDaBal}. Since that strong increase in $D\alpha$ emission is not captured by the 'default' setup, this causes strong concerns on the synthetic deuterium (and tritium) atomic emission diagnostic signals predicted from plasma-edge modelling. This has implications on diagnostic design, analysis of spectroscopic diagnostics as well as real-time control strategies that use spectroscopic signals as a real-time sensor. 

Synthetic diagnostic signals of hydrogen emission are used to test spectroscopic analysis techniques \cite{Bowman2020,Perek2022,Verhaegh2019a,Verhaegh2021b} and design diagnostics \cite{Kukushkin2016a}. For example, synthetic diagnostics have shown that unexpectedly high stray light emission from hydrogen, deuterium and tritium Balmer-$\alpha$ emission can be a concern for diagnostic interpretation in ITER \cite{Kukushkin2016a}. Plasma-molecular chemistry with molecular ions could, if present, greatly enhance the divertor hydrogenic emission beyond that predicted in the simulations; which could grossly misinform studies relying on synthetic diagnostic data.

Plasma-molecular effects result in a significantly enhanced population of the hydrogen atom $n=3$ state, which may have implications for the treatment of photon opacity to the Lyman series in simulations \cite{Pshenov2019,Pshenov2023} as well as the diagnosis of photon opacity \cite{Verhaegh2021b}. Accounting for molecular ions in the analysis of hydrogen atomic line emission required the creation of novel analysis techniques \cite{Karhunen2022a,Karhunen2022,Verhaegh2021a}, which need to be further expanded to include photon opacity effects. Additionally, the combination of plasma-molecular interactions and photon opacity will, even more greatly, amplify the $D\alpha$ emission as photon opacity of Lyman-$\beta$ emission results in more $D\alpha$ emission.

Real-time detachment control strategies are required in reactors and spectrally filtered imaging \cite{Ravensbergen2021,Ravensbergen2020} as well as line-of-sight passive spectroscopy of hydrogen atomic emission \cite{Biel2019} are important detachment sensor candidates. The complexity of including molecular ions in the interpretation of the atomic hydrogenic emission as well as the occurrence of photon opacity can complicate the usage of hydrogen atomic emission for such purposes and using complementary or alternative methods such as monitoring the molecular Fulcher band intensity may be required \cite{Verhaegh2022}. Plasma-edge simulations with an improved plasma-molecular interaction set as well as photon opacity are required to investigate this further.

\subsection{Inaccuracies of the molecular charge exchange rate employed by Eirene}
\label{ch:rates_improvement}

Increasing the molecular charge exchange rate during detachment through modified rates is a first step in 1) explaining the discrepancy observed between the experiment and SOLPS-ITER simulation results for TCV in terms of MAR, hydrogen emission and the ion target flux; 2) the investigation of the importance of molecular ions during detachment. This work aims to motivate that a rigorous revision and re-derivation of the various molecular rates in plasma-edge codes is required and below we will discuss the three inaccuracies of the molecular charge exchange rate used by Eirene, introduced in section \ref{ch:molCX}, in more detail.

First, we investigate the inaccuracies of the vibrationally resolved molecular charge exchange cross-sections and their impact. As explained in section \ref{ch:molCX}, the cross-sections are underestimated at low velocities for higher vibrational levels as a simplified equation \cite{Greenland2001} is used to rescale the measured cross-sections from the vibrational ground state \cite{Janev1987,Holliday1971} to higher vibrational levels (referred to as 'Janev 1987 / Holliday 1971 / Greenland 2001'): leading to greatly underestimated effective rates at low temperatures. This is in strong contrast to more recent, fully vibrationally resolved, calculations of the molecular charge exchange cross-sections \cite{Ichihara2000,Laporta2021, Krstic2003,Roncero2022}, referred to as 'Ichihara 2000' \cite{Ichihara2000}. The underestimate of the effective rates at low temperatures are exacerbated by ion isotope mass rescaling for deuterium (equation \ref{eq:eirene_rate}) and even more so for tritium. 

\begin{figure}
    \centering
    \includegraphics[width=0.75\linewidth]{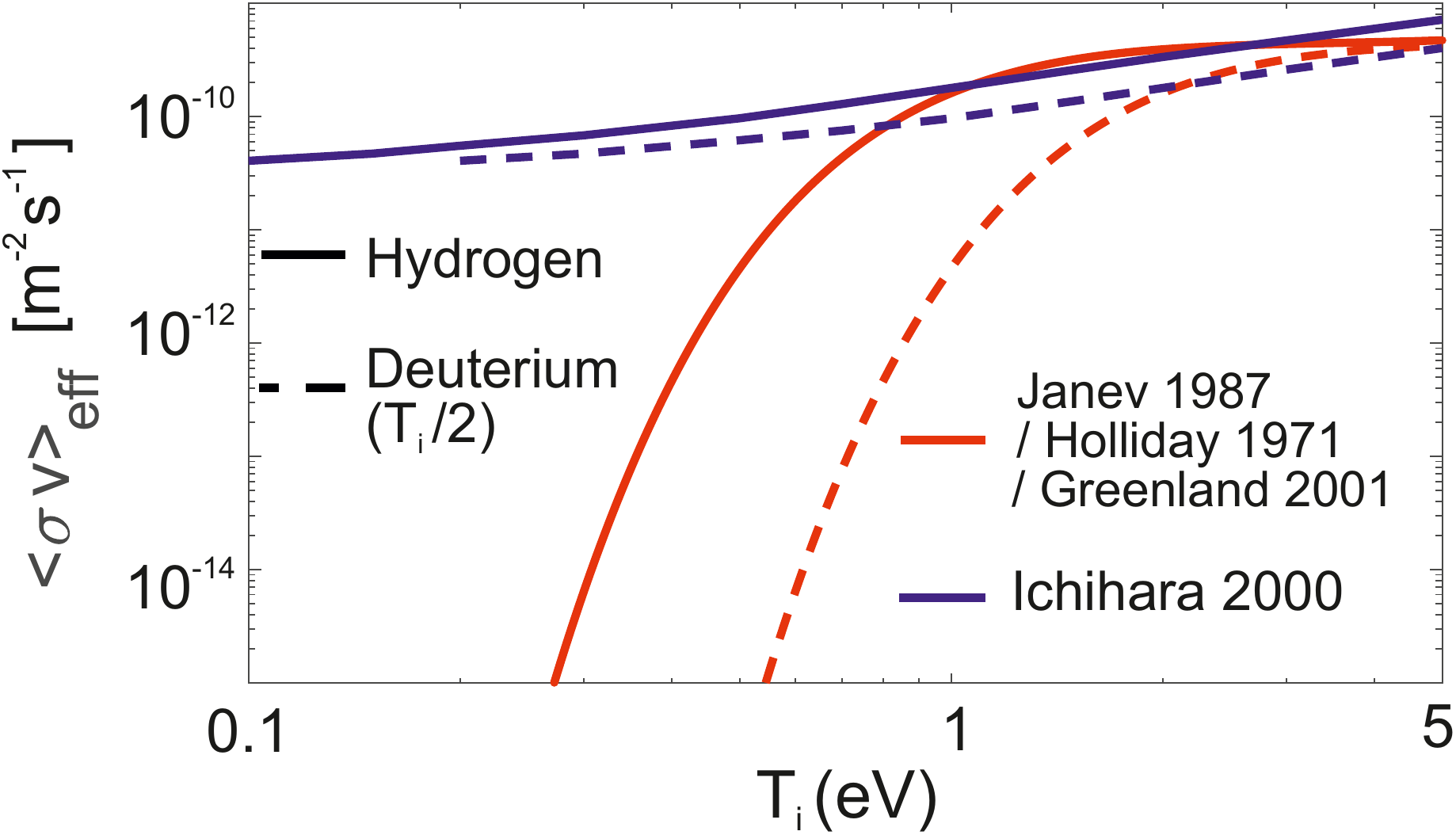}
    \caption{Comparison of the effective molecular charge exchange rate, as function of the ion temperature, using the vibrationally resolved molecular charge exchange rates from Holliday 1971, Janev 1987, Greenland 2001 (red) \cite{AMJUEL, Janev1987, Holliday1971,Greenland2001} and from Ichihara 2000 \cite{Ichihara2000} (blue). The effective rate is calculated using equation \ref{eq:eff_rate}. The vibrational distribution is obtained from \cite{Holm2022}, which has been averaged over $T_e =1 - 3$ eV. Static molecules ($E_{H_2} = 0.1$ eV) has been assumed. Purely hydrogenic rates have been used and ion isotope mass rescaling ($T_i / 2$) has been applied to the dotted cases.}
    \label{fig:MolCX}
\end{figure}

The impact of this on the effective molecular charge exchange rate is investigated in figure \ref{fig:MolCX}, where the effective molecular charge exchange rate is calculated as function of $T_i$ using a fixed vibrational distribution (obtained from \cite{Holm2022} assuming $T_e = 1-3$ eV) for both vibrationally resolved molecular charge exchange cross-sections \footnote{A fixed $E_{H_2}$ can introduce uncertainties since higher molecular energies can elevate the effective cross-sections at low ion temperatures significantly. Nevertheless, this would not alter the conclusions of figure \ref{fig:MolCX}.}. This shows that, although the ion isotope mass rescaling is correctly applied to only the ion temperature dependency, the effective cross-section greatly decays at low temperature for 'Janev 1987 / Holliday 1971 / Greenland 2001' for D. Contrastingly, the effective rates derived using the 'Ichihara 2000' cross-sections are similar for H and D, in agreement with \cite{Reiter2018}, and are both in reasonable agreement with the 'Janev 1987 / Holliday 1971 / Greenland 2001' effective rate for H $T_e > 2$ eV. 

Secondly, as explained through equations \ref{eq:eff_rate} and \ref{eq:eirene_rate} in section \ref{ch:molCX}, both the ion and electron temperature dependencies are inadvertently rescaled when ion isotope mass rescaling is applied by Eirene to effective rates. This results in inaccuracies \emph{not only} because it results in an incorrectly applied ion isotope mass rescaling to the electron temperature dependency of the vibrational distribution model; \emph{but also} because the ion temperature can be different from the electron temperature. Resolving this may require modifications to Eirene to support different electron and ion temperature dependencies. Using a vibrationally resolved simulation setup (see section \ref{ch:vibr_distr}) would also ensure that ion mass rescaling is applied correctly.

Thirdly, as mentioned in section \ref{ch:molCX}, it is assumed that the cross-sections are the same for all isotopes. This is not true, however, as there are chemical differences resulting in different cross-sections for each isotope \cite{Ichihara2000, Reiter2018}. The chemical isotope differences have a particularly strong impact on the rates resulting in $D^-$. 

SOLPS-ITER does not account for $H^-$ by default. In \cite{Kukushkin2017} it was argued that such interactions can play an important role as they also result in MAR; that argument was based on applying the correct ion isotope mass re-scaling under the assumption that the cross-sections for creating $H^-$ is the same as for $D^-$ and $T^-$. However, the $D^-$ and $T^-$ creation cross-sections are strongly reduced compared to the $H^-$ ones due to chemical isotopical differences in the various rates, which has been measured experimentally \cite{Krishnakumar2011}. Such measurements, however, occur at very low vibrational levels. The isotope differences are expected to reduce at higher vibrational levels, which drive most of the molecular ion generation \cite{Reiter2018}. As such, a more detailed analysis in \cite{Reiter2018} indicates a 30 \% reduction in the effective $D^-$ creation rate compared to $H^-$. However, that percentage, as well as the $H^-$ generation rate, will be even more sensitive to the vibrational distribution. This may imply that it is necessary, in some conditions, to include interactions with $H^-$ and its isotopes in plasma-edge modelling. There are experimental indications from the TCV tokamak on the presence of $D^-$ during deeply detached conditions, based on the inferred ratio between the 'molecular' contribution to $D\beta$ and $D\alpha$ \cite{Verhaegh2021b}.

Plasma interactions with $D_2^+$ did not seem to contribute significantly to the volumetric momentum losses, even for the 'modified' rate setup. Since the momentum loss from plasma-molecular interactions is similar for both reaction setups, despite the orders-of-magnitude differences in the molecular charge exchange rates, molecular charge exchange may not be contributing significantly to the momentum balance. However, since $D_2^+$ is a 'test ion' that is static in Eirene after it is created, there may be inaccuracies in how momentum balance is accounted for in SOLPS-ITER when significant amounts of $D_2^+$ are generated, which needs further investigation.

\subsection{Uncertainties in the vibrational distribution of molecular hydrogen}
\label{ch:vibr_distr}

Molecular charge exchange is highly sensitive to the vibrational distribution of the molecule at the time of the reaction ($f_\nu$). The modelling of $f_\nu$ has large uncertainties, which can be broadly divided in two categories: 1) inaccuracies in the rates and reactions used in the vibrational distribution modelling; 2) inaccuracies introduced by a lack of transport. 

The first category includes inaccuracies in \emph{reaction rates} used as well as missing \emph{reaction processes}, including 1) the omission of $H^-$ creation; 2)  re-distribution of vibrationally excited states through electronic excitation \cite{Chandra2023, Holm2022}; 3) omitting electron-impact collisions that alter the vibrational state of a molecule by more than $> \pm 1$ \cite{Holm2022}. Including the latter two in the vibrational modelling can alter the vibrational distribution considerably \cite{Holm2022}.

The mean-free-path of vibrationally excited molecules can be sufficiently long for transport to be significant, particularly below the dissociation region. Including such transport requires vibrationally resolved simulations \cite{Fantz2001,Fantz2002,Fantz2006, Wischmeier2005}. The vibrational distribution can vary strongly spatially and transport allows including such effects and their propagation throughout the rest of the divertor. Plasma-wall interactions \cite{Wischmeier2005} can alter the initial vibrational distribution of molecules coming off the wall, depending on the precise interaction with the wall and the wall material \cite{Wischmeier2004,Eenhuistra1988}.

Although vibrationally resolved simulations have been performed in the past for Asdex-Upgrade \cite{Fantz2001,Fantz2002} and for TCV \cite{Wischmeier2004, Wischmeier2005}, they may not have included all the relevant processes (e.g. inaccuracies in the rates \& reactions of vibrationally excited molecules \cite{Holm2022}) and would have employed the default Eirene cross-sections that are likely strongly underestimated at high vibrational levels (section \ref{ch:rates_improvement}). Therefore, molecular charge exchange in detached conditions was, likely, still significantly underestimated in these simulations. Further investigation of the vibrational distribution of the molecules, through both modelling and experiment to modelling comparisons, is required in conditions where plasma chemistry with molecular ions may be important.

\section{Conclusions}
\label{ch:conclusion} 

Recent experimental results on TCV, MAST-Upgrade and JET have indicated that plasma-molecular chemistry, resulting in molecular ions (particularly $D_2^+$) that react with the plasma, result in excited atoms that can contribute to the hydrogen Balmer line emission significantly. Such interactions result in Molecular Activated Recombination (MAR), which can impact divertor particle balance significantly during detachment. Initial comparisons between SOLPS-ITER simulations and experiments on TCV had shown that such interactions do not occur significantly in the simulations. It was hypothesised that this is related to the isotope mass rescaling employed by Eirene to the effective hydrogenic molecular charge exchange rate, resulting in $\sim 100$ times lower $D_2^+$ densities in detachment-relevant regimes compared to $H_2^+$, whereas more detailed investigations in literature indicate differences between hydrogen and deuterium molecular charge exchange rates of a few percent.

This has motivated our work to compare SOLPS-ITER simulation results with the default rate setup and with a modified rate setup in which ion isotope mass rescaling has been disabled for molecular charge exchange. We observe that disabling isotope mass rescaling for molecular charge exchange has a strong impact on the solution obtained after the detachment onset and provides a closer match to the experiment.  The neutral atom content in the lower divertor is greatly enhanced in the modified rate setup by up to 100 \%, due to Molecular Activated Dissociation (MAD) and MAR, which has significant associated hydrogenic (plasma) power losses. 

\section{Acknowledgements}

Discussions with Detlev Reiter are kindly acknowledged and were very helpful. This work has received support from EPSRC Grants EP/T012250/1 and EP/N023846/1. This work has been supported in part by the Swiss National Science Foundation and has been carried out within the framework of the EUROfusion Consortium, funded by the European Union via the Euratom Research and Training Programme (Grant Agreement No 101052200 — EUROfusion; as well as No 633053 (2014-2018 \& 2019-2020)). Views and opinions expressed are however those of the author(s) only and do not necessarily reflect those of the European Union or the European Commission. Neither the European Union nor the European Commission can be held responsible for them.

\section{References}

\bibliographystyle{iopart-num}
\bibliography{all_bib.bib}

\end{document}